%% file: draft.tex
\documentclass[final,5p,twocolumn, times]{elsarticle}
\usepackage{amssymb}
\usepackage{graphicx}
\usepackage{geometry}
\usepackage[dvipsnames,table]{xcolor}
\usepackage{slashed}
\usepackage{hyperref} 
\usepackage{xspace}
\usepackage[tight]{subfigure}
\usepackage{amsmath}
\usepackage{amsfonts}
\usepackage{mathrsfs}
\usepackage{comment}
\usepackage{afterpage}
\usepackage{verbatim}
\usepackage{booktabs}
\usepackage{array}
\usepackage[title,titletoc]{appendix}
\usepackage{tabularx}
\newcolumntype{C}[1]{>{\centering\arraybackslash}p{#1}}

\newcommand{\mc}{\mathcal}

\input{macros.tex}

\journal{Physics Letters B}

\begin{document}
\begin{frontmatter} 
  \title{
    NLO QCD predictions for doubly-polarized $\PW\PZ$ production at the LHC
  }
  \author[label1]{Ansgar Denner}\ead{ansgar.denner@physik.uni-wuerzburg.de}
  \author[label1]{Giovanni Pelliccioli}\ead{giovanni.pelliccioli@physik.uni-wuerzburg.de}
  \address[label1]{University of W\"urzburg, Instit\"ut f\"ur Theoretische Physik und Astrophysik,
  Emil-Hilb-Weg 22, 97074 W\"urzburg (Germany)}
\begin{abstract}
   Accessing the polarization of weak bosons provides an important probe for the
   mechanism of electroweak symmetry breaking. Relying on the double-pole approximation
   and on the separation of polarizations at the amplitude level,
   we study $\PW\PZ$ production at the LHC, with both bosons in a definite
   polarization mode, including NLO QCD effects. We compare results obtained
   defining the polarization vectors in two different frames. Integrated and differential
   cross-sections in a realistic fiducial region are presented.
\end{abstract}
\end{frontmatter}

\section{Introduction}\label{intro}
The investigation of di-boson production represents a crucial target for the
Large Hadron Collider (LHC) with the accumulated Run 2 data.
These processes provide a direct probe of the non-abelian character
of electroweak interactions in the Standard Model (SM) as well as of
possible deviations from
standard triple-gauge-boson-coupling interactions due to new physics. On top of this, the
study of the weak-boson polarization supplies a probe of the mechanism
of electroweak symmetry breaking, which is responsible for the
appearance of a longitudinal
polarization mode of massive gauge bosons.

Among di-boson processes, $\PW^\pm\PZ$ production with fully-leptonic
decays is optimal to
extract detailed information on polarizations, as it features a sizeable rate at the LHC
(higher than $\PZ\PZ$, smaller than $\PW^+\PW^-$) and gives almost complete
access to the final-state kinematics, up to the
reconstruction of the single neutrino (impossible in fully-leptonic $\PW^+\PW^-$).

The CMS and ATLAS collaborations have measured the inclusive and differential
production of $\PW^\pm\PZ$ at the LHC at $13\TeV$ centre-of-mass (CM) energy
\cite{Khachatryan:2016tgp,Aaboud:2016yus,Sirunyan:2019bez}, and recently
ATLAS has also measured polarization fractions for the $\PW$ and the $\PZ$ boson
in this process \cite{Aaboud:2019gxl}.
In the experimental community there is increasing interest in the measurement
of the polarization of single bosons produced in electroweak processes, as well
as in the extraction of signals with two polarized bosons, in processes like
inclusive di-boson production and vector-boson scattering.

The $\PW\PZ$ hadronic production in the SM has been intensely investigated
in the theoretical community. The next-to-leading-order (NLO) QCD corrections are known since
many years \cite{Ohnemus:1991gb, Frixione:1992pj}. NLO electroweak (EW) corrections
have been calculated for the on-shell \cite{Bierweiler:2013dja,Baglio:2013toa}
and the off-shell case \cite{Biedermann:2017oae}. Next-to-next-to-leading-order (NNLO)
QCD corrections are known in the on-shell case \cite{Grazzini:2016swo}, including
off-shell effects \cite{Grazzini:2017ckn} and combined with EW corrections
\cite{Kallweit:2019zez}. Beyond fixed order, computations including resummed predictions
\cite{Becher:2019bnm} and matching to parton shower \cite{Nason:2013ydw,Chiesa:2020ttl}
are also available. $\PW\PZ$ production is also relevant for the study of anomalous
triple-gauge-boson couplings \cite{Baur:1994aj,Franceschini:2017xkh,Chiesa:2018lcs,Baglio:2019uty}.

Weak-boson polarizations in $\PW\PZ$ production at the LHC
have been first investigated in \citere{Stirling:2012zt} at leading order (LO).
NLO QCD and EW effects on fiducial polarization observables have been studied in
\citeres{Baglio:2018rcu,Baglio:2019nmc} in the case where only one of the two
bosons has a definite polarization state.
The automated Monte Carlo simulation of  boson-pair production with polarized weak bosons
is currently well-established in \madgraph 5 \cite{BuarqueFranzosi:2019boy} in the
narrow-width approximation (NWA), but the perturbative accuracy is limited to LO.
Since the measurement of $\PW\PZ$ production with both bosons in a definite
polarization mode (which we call \emph{doubly-polarized} for simplicity) represents
a major aim for experimental collaborations, an improved theoretical description of these
polarized signals is definitely needed to compare with data.

The work presented in this paper, using the methods introduced in Ref.~\cite{Denner:2020bcz},
improves the perturbative accuracy for doubly-polarized $\PW\PZ$ production at the LHC,
including NLO QCD effects, and enables a precise modelling of off-shell effects and
interferences among polarized amplitudes.

Furthermore, recent experimental results \cite{Aaboud:2019gxl,CMS:2020zoz}
and theoretical studies \cite{Baglio:2019nmc,Ballestrero:2020qgv} in di-boson  
and vector-boson scattering have addressed a novel definition of polarizations
in the \emph{modified helicity coordinate system}. In this definition,
the reference axis is fixed by the direction of the corresponding
boson in the di-boson CM frame,
and not in the laboratory frame (\emph{helicity coordinate system} \cite{Bern:2011ie}).
We present in this paper results for both polarization definitions, analysing the
main differences under a theoretical perspective and keeping in mind the limited
sensitivity of the LHC to polarized signals.
Even though some results for singly-polarized cross-sections are shown, the main focus
of this paper is put on doubly-polarized predictions.

The paper is organized as follows. In Sect.~\ref{details} we briefly recall the
methods employed to define a polarized signal at the amplitude level.
We also detail the SM parameters and selection cuts that we have used in Monte Carlo
simulations. In Sect.~\ref{results} we show numerical results for polarized signals
with both definitions of polarizations. Our conclusions are given in Sect.~\ref{concl}.

\section{Details of the calculation}\label{details}

\subsection{Polarized signal definition}\label{subsec:calcdetails}
In order to define singly- and doubly-polarized signals in $\PW\PZ$ production,
we apply the technique detailed in \citere{Denner:2020bcz}, which relies on the
double-pole approximation (DPA) \cite{Aeppli:1993cb,Aeppli:1993rs,Beenakker:1998gr,
  Denner:2000bj,Billoni:2013aba,Biedermann:2016guo}
for the gauge-invariant treatment of resonant diagrams, and on the separation of
polarizations in weak-boson propagators at the amplitude level \cite{Ballestrero:2017bxn}.
This technique has been applied also to real-radiation squared amplitudes
and to subtraction counterterms that contribute to the NLO QCD cross-sections.
This ensures that a specific polarization mode for one or both bosons is coherently
selected in all parts of the calculation. Such a strategy allows us to
define polarized signals in a natural manner, enabling the modelling of off-shell effects and spin
correlations. Moreover, it is expected to be more accurate than methods that rely
either on the NWA or on a reweighting procedure \cite{Ballestrero:2019qoy}.
Furthermore, defining polarizations at the amplitude level enables to account for
the correlation between the two weak bosons, which are very important in di-boson
production \cite{Denner:2020bcz}.

One purpose of this paper is to investigate different polarization definitions.
It is well known that the polarization vectors for a weak boson need to be defined
in a specific reference frame. In di-boson production, natural frames are the
laboratory (LAB) and the di-boson centre-of-mass (CM) frame. In the definition of
\citere{Bern:2011ie}, the reference axis for polarization vectors of a single
boson is fixed by its direction in the LAB reference frame. In the one of 
\citere{Aaboud:2019gxl}, the reference axis is the corresponding direction in the
di-boson CM frame. Results in the two definitions cannot be easily related
to each other as the two reference frames are connected by non-trivial Lorentz
transformations. However, a comparison is useful to extract information
about the spin structure of the process, as well as to highlight relevant observables
which maximize the discrimination power among polarization states.

In the following, two polarization modes are considered for weak bosons: longitudinal (L)
and transverse (T). The transverse one is understood as the coherent sum of the left-
and right-handed modes. This choice results in four doubly-po\-la\-rized signals:
$\PW_{\rL}\PZ_{\rL},\,\PW_{\rL}\PZ_{\rT},\,\PW_{\rT}\PZ_{\rL},\,\PW_{\rT}\PZ_{\rT}$.
Singly-po\-la\-rized signals for the $\PW(\PZ)$ boson are labelled as $\PW_{\lambda}\PZ_{\rU}$
($\PW_{\rU}\PZ_{\lambda}$), where $\lambda=\rm L,T$, and $\rU$ stands for unpolarized.

As in \citere{Denner:2020bcz}, the polarized and unpolarized tree-level and one-loop
amplitudes are provided by \recola \cite{Actis:2012qn, Actis:2016mpe}
and \collier \cite{Denner:2014gla,Denner:2016kdg}. The numerical
integration is performed with the \mocanlo Monte Carlo code, which embeds
Catani--Seymour dipoles \cite{Catani:1996vz} for the subtraction of infrared singularities,
and is interfaced with parton distribution functions (PDFs) via {\scshape{LHAPDF6}
  \cite{Buckley:2014ana}}.

\subsection{Setup(s)}\label{subsec:setup}

We consider the process $\Pp \Pp \rightarrow \Pe^+\nu_{\Pe}\,\mu^+\mu^- + X$
at $13\TeV$ CM energy.
At tree-level [$\mc O (\alpha^4)$], such a process receives contributions only from
the $q\bar{q}$ partonic channel, while the inclusion of QCD radiative corrections
[$\mc O (\alphas \alpha^4)$] enables real contributions from the $q\Pg(\bar{q}\Pg)$ initial
state.

Following Ref.~\cite{Tanabashi:2018oca}, the weak-boson on-shell masses and widths
are chosen as
\begin{align}
\MW^{\rm os} ={}& 80.379\GeV\,,&  \GW^{\rm os}   ={}& 2.085\GeV\,, \notag \\
\MZ^{\rm os} ={}& 91.1876\GeV\,,&   \GZ^{\rm os} ={}& 2.4952\GeV\,,
\end{align}
and then converted into the corresponding pole masses \cite{Bardin:1988xt} which serve
as inputs for the Monte Carlo simulations. The electroweak coupling is fixed in the
$G_\mu$~scheme, with
\beq
\GF = 1.16638\cdot10^{-5} \GeV^{-2}\,.
\eeq
The Higgs-boson and top-quark parameters do not enter the computation at the considered accuracy.
Unstable bosons and weak couplings are treated in the complex-mass scheme
\cite{Denner:2000bj,Denner:2005fg,Denner:2006ic}.
We employ \textsc{NNPDF3.1} parton distribution functions \cite{Ball:2017nwa}, computed with
$\alphas(\MZ)=0.118$ at LO and NLO for the LO and NLO QCD predictions, respectively.
The renormalization and factorization scales are simultaneously set to
$\mu_{\rF}=\mu_{\rR}=(\MZ+\MW)/2$ (pole masses).

To validate polarized calculations, we have studied an \emph{inclusive setup},
which only includes a cut on the $\mu^+\mu^-$ invariant mass that is constrained
to be close to the $\PZ$ mass,
\beq\label{eq:mllcut}
81\GeV<M_{\mu^+\mu^-}<101\GeV\,.
\eeq
Then, we have considered a \emph{fiducial setup} that mimics the signal region of
\citere{Aaboud:2019gxl}. The following selections are applied:
\begin{itemize}
\item a minimum transverse momentum cut for the positron, $\pt{\Pe^+}>20\GeV$, and for
  the (anti)muon, $\pt{\mu^\pm}>15\GeV$;
\item a maximum rapidity cut for all charged leptons, $|y_\ell|<2.5$;
\item a minimum rapidity--azimuthal-angle distance cut for lepton
  pairs, $\Delta R_{\mu^+\mu^-}>0.2$ and $\Delta R_{\mu^\pm\Pe^+}>0.3$;
\item a lower cut on the $\PW$-boson transverse mass,
  \beq
  \mt{\PW} = \sqrt{2\,\pt{\Pe^+}\pt{\nu_{\Pe}}\left(1-\cos\Delta\phi_{\Pe^+\nu_{\Pe}}\right)} > 30\GeV\,.
  \eeq
\item the cut specified in Eq.~\refeq{eq:mllcut}.
\end{itemize}
No veto is imposed on the hadronic jets that may originate from real radiation at NLO QCD.

We have only considered the case of two unequal-flavour pairs of leptons. Performing
an analogous study for same-flavour leptons would not introduce further
conceptual issues in the signal definition. In fact, we expect that the presence
of identical leptons would not lead to large interference effects. Furthermore,
the additional ambiguity in identifying the candidate leptons reconstructing the
$\PZ$ and the $\PW$ boson can be treated by proper selection cuts.

All of the presented results concern the production of a $\PW^+\PZ$ in the
\emph{fiducial setup}. The charge-conjugate process can be studied analogously.

\begin{table*}[h]
\begin{center}
\renewcommand{\arraystretch}{1.3}
\begin{tabular}{C{3.8cm}C{2.3cm}C{1.7cm}C{2.3cm}C{1.7cm}C{1.7cm}}%
\hline %
\cellcolor{blue!14}   & \cellcolor{blue!14}  $\sigma_{\rm LO}$ [fb]  & \cellcolor{blue!14} fraction$_{\rm LO}$& \cellcolor{blue!14} $\sigma_{\rm NLO}$ [fb] & \cellcolor{blue!14} fraction$_{\rm NLO}$ & \cellcolor{blue!14} {$K$-factor}      \\
\hline
full   &   $ 19.537(7)^{+ 4.1 \%}_{- 4.9 \%} $ & - & $  35.27(1)^{+ 5.2 \%}_{- 4.2 \%} $  & - &  1.81   \\
unpolarized (DPA)   &   $ 19.125(6)^{+ 4.1 \%}_{- 5.0 \%} $ & 100\%  & $  34.63(1)^{+ 5.3 \%}_{- 4.2 \%} $  & 100\% &   1.81   \\[.5ex]
\hline
\multicolumn{6}{c}{\cellcolor{green!9} polarizations defined in the CM frame} \\
\hline
$\PW^+_{\rL}\PZ^{\,}_{\rU}$ (DPA)   &   $ 3.502(1)^{+ 4.8 \%}_{- 5.7 \%} $ & 18.3\%&  $  7.308 ( 2 )^{+ 6.1 \%}_{- 4.9 \%} $  & 21.1\%&   2.09   \\
$\PW^+_{\rT}\PZ^{\,}_{\rU}$ (DPA)   &   $ 15.480(6)^{+ 4.0 \%}_{- 4.8 \%} $ & 80.9\%&  $  27.14 ( 1 )^{+ 5.0 \%}_{- 4.1 \%} $  &78.4\%  &   1.75   \\[.5ex]
\hline
$\PW^+_{\rU}\PZ^{\,}_{\rL}$ (DPA)   &   $ 3.451(1)^{+ 4.8 \%}_{- 5.7 \%} $ & 18.0\%&  $  7.137 ( 2 )^{+ 6.0 \%}_{- 4.9 \%} $  & 20.6\%&   2.07   \\
$\PW^+_{\rU}\PZ^{\,}_{\rT}$ (DPA)   &   $ 15.674(6)^{+ 4.0 \%}_{- 4.8 \%} $ & 81.9\%&  $  27.449 ( 9 )^{+ 5.0 \%}_{- 4.1 \%} $  & 79.2\%&    1.75   \\[.5ex]
\hline
$\PW^+_{\rL}\PZ^{\,}_{\rL}$ (DPA)   &   $ 1.508 ( 1 )^{+ 4.5 \%}_{- 5.3 \%} $ & 7.9\%&  $  1.968 ( 1 )^{+ 2.7 \%}_{- 2.2 \%} $  &  5.7\%&   1.31   \\
$\PW^+_{\rL}\PZ^{\,}_{\rT}$ (DPA)   &   $ 2.018 ( 1 )^{+ 5.1 \%}_{- 6.0 \%} $ & 10.6\%&  $  5.354 ( 1 )^{+ 7.3 \%}_{- 5.9 \%} $  & 15.5\% &   2.65   \\
$\PW^+_{\rT}\PZ^{\,}_{\rL}$ (DPA)   &   $ 1.902 ( 1 )^{+ 5.0 \%}_{- 5.9 \%} $ & 9.9\%&  $  5.097 ( 2 )^{+ 7.4 \%}_{- 5.9 \%} $  & 14.7\% &   2.68   \\
$\PW^+_{\rT}\PZ^{\,}_{\rT}$ (DPA)   &   $ 13.555 ( 5 )^{+ 3.8 \%}_{- 4.7 \%} $ & 70.9\%&  $  21.992 ( 9 )^{+ 4.5 \%}_{- 3.6 \%} $  & 63.5\%&     1.62   \\[.5ex]
\hline
\multicolumn{6}{c}{\cellcolor{green!9} polarizations defined in the LAB frame} \\
\hline
$\PW^+_{\rL}\PZ^{\,}_{\rU}$ (DPA)   &   $ 4.227(1)^{+ 4.8 \%}_{- 5.7 \%} $ & 22.1\%& $  8.160 ( 2 )^{+ 5.6 \%}_{- 4.6 \%} $  & 23.6\%&   1.93   \\
$\PW^+_{\rT}\PZ^{\,}_{\rU}$ (DPA)   &   $ 14.865(5)^{+ 3.9 \%}_{- 4.8 \%} $ & 77.7\% &  $  26.394 ( 9 )^{+ 5.1 \%}_{- 4.1 \%} $  & 76.2\%&   1.78   \\[.5ex]
\hline
$\PW^+_{\rU}\PZ^{\,}_{\rL}$ (DPA)   &   $ 5.528 (2 )^{+ 4.7 \%}_{- 5.6 \%} $ & 28.9\%& $  9.550 ( 4 )^{+ 4.9 \%}_{- 4.0 \%} $  &27.6\% &   1.73   \\
$\PW^+_{\rU}\PZ^{\,}_{\rT}$ (DPA)   &   $ 13.654 (5 )^{+ 3.9 \%}_{- 4.8 \%} $ & 71.4\%& $  25.052 (8 )^{+ 5.3 \%}_{- 4.3 \%} $  & 72.3\%&   1.83   \\[.5ex]
\hline
$\PW^+_{\rL}\PZ^{\,}_{\rL}$ (DPA)   &   $ 1.0824 ( 4 )^{+ 4.1 \%}_{- 4.9 \%} $ & 5.7\%&   $  2.063 ( 1 )^{+ 5.6 \%}_{- 4.5 \%} $  & 6.0\% &    1.91   \\
$\PW^+_{\rL}\PZ^{\,}_{\rT}$ (DPA)   &   $ 3.165 ( 1 )^{+ 5.1 \%}_{- 6.0 \%} $ &  16.5\%&  $  6.108 ( 2 )^{+ 5.6 \%}_{- 4.5 \%} $  & 17.6\% &   1.93   \\
$\PW^+_{\rT}\PZ^{\,}_{\rL}$ (DPA)   &   $ 4.381 ( 2 )^{+ 4.8 \%}_{- 5.7 \%} $ & 22.9\% &  $  7.409 ( 4 )^{+ 4.8 \%}_{- 3.8 \%} $  & 21.4\% &   1.69   \\
$\PW^+_{\rT}\PZ^{\,}_{\rT}$ (DPA)   &   $ 10.526 ( 4 )^{+ 3.6 \%}_{- 4.4 \%} $ & 55.0\%&   $  18.964 (7 )^{+ 5.2 \%}_{- 4.2 \%} $  &54.8\% &    1.80   \\[.5ex]
\hline
\end{tabular}
\end{center}
\caption{Integrated cross-sections (in fb) in the fiducial setup for unpolarized,
  singly-polarized and doubly-polarized $\PW^+\PZ$ production. Polarizations
  are defined in the CM and the LAB frames. Numerical errors (in parentheses) and
  uncertainties from 7-point scale variations (in percentage) are shown. The
  fractions (in percentage) are computed as ratios of polarized cross-sections over the
  DPA unpolarized one. $K$-factors are computed as ratios of the NLO QCD over the LO
  cross-sections.} 
\label{table:sigmainclNLO}
\end{table*}

\section{Results}\label{results}
We have thoroughly validated our calculation in the inclusive setup introduced
in Sect.~\ref{subsec:setup}.
In the absence of cuts on the leptons, interferences
among polarization states vanish for most of the relevant variables, and
polarized signals can be extracted by suitable projections on
the unpolarized distributions in the decay angles
\cite{Stirling:2012zt,Ballestrero:2017bxn}. The comparison of Monte
Carlo results for polarized cross-sections
with those projected out of the unpolarized distributions has given perfect agreement
both for polarization fractions and in the shapes of decay-angle
distributions ($<0.5\%$ discrepancy at both integrated and differential level).
This holds for singly-polarized signals, but also for doubly-polarized results
that can be analogously extracted from the singly-polarized ones. The validation
has been done both for polarizations defined in the CM and in the LAB frames. 

As a further check, we have compared LO integrated polarized results in the CM
definition with those obtained with \madgraph 5 \cite{BuarqueFranzosi:2019boy},
finding good agreement ($\lesssim 1\%$ level). 

In the following we present selected integrated and differential
results for the fiducial setup described in Sect.~\ref{subsec:setup}.

\subsection{Integrated cross-sections}\label{subsec:integrated}
Very large $K$-factors characterize the production of $\PW\PZ$
at high-energy hadron collisions. The large QCD corrections
are related to the approximate radiation-zero at Born level
\cite{Baur:1994ia}, which suppresses the LO prediction {but
is not} present in the real corrections at NLO QCD
\cite{Grazzini:2016swo}.

In Table~\ref{table:sigmainclNLO} we show integrated
cross-sections in the fiducial region for unpolarized and
polarized bosons. 
%
The full process is considered as a reference, as it includes
all resonant and non-resonant effects. The unpolarized process
in DPA is regarded as the unpolarized di-boson signal.
The difference between the full and the DPA
unpolarized results is considered as a background to di-boson
production \cite{Denner:2020bcz}, and amounts to 2.1\% (1.8\%)
at LO (NLO) QCD at the integrated level. 

Summing singly- or doubly-polarized cross-sections
and comparing the result against the DPA unpolarized result,
one can extract information about the interferences among
polarization states. Surprisingly, these effects never exceed $0.8\%$ in absolute value,
with both the LAB and the CM definition, despite the
application of lepton cuts which, in general, are expected
to generate non-vanishing interferences \cite{Belyaev:2013nla,Stirling:2012zt}.

The cross-sections for a polarized $\PW^+$ and an unpolarized $\PZ$~boson show similar
behaviours in the two definitions: the longitudinal $K$-factor is
about $2$, the longitudinal fraction is roughly 21\% in the CM
definition, not so far from the 24\% in the LAB definition. 
A slightly different situation is found for a polarized $\PZ$ boson
produced with an unpolarized $\PW$, likely due to the asymmetric cuts on the
leptons: the longitudinal fraction in the LAB definition (28\%) is higher than in
the CM one (21\%), despite a larger impact of QCD corrections in the CM one. For
singly-polarized results in both definitions, the fractions are only mildly
sensitive to the inclusion of QCD corrections.

The differences between the CM and the LAB definitions of
polarizations are much more evident when
considering doubly-polarized cross-sections.
The polarization fractions in the LAB definition are roughly conserved when going
from LO to NLO QCD, as already found in $\PW^+\PW^-$ production \cite{Denner:2020bcz}, while
this is not the case in the CM definition. Furthermore, very different $K$-factors emerge
in the two definitions: in the CM one both mixed polarization
states feature a +170\% enhancement due to QCD corrections, while in the
LAB one the LT and TL polarization states are enhanced by 90\% and 70\%, respectively.
The LL polarization mode defined in the CM frame features a very small $K$-factor (1.31), which is surprising
given the larger QCD corrections in the singly-longitudinal cases. On the contrary,
the same mode in the LAB definition features a +90\% enhancement due to QCD corrections.
At NLO QCD, the LL signal amounts to roughly 6\% of the total
in both definitions. The TT signal is larger in the CM definition, and conversely the
mixed contributions are larger in the LAB one.


{The back-to-back kinematics of the bosons in the CM frame at LO implies that the
polarization vectors of the two weak bosons are defined with respect to the same
reference axis, up to a change of sign. This leads to very similar results for the LT
and TL polarization modes defined in the CM frame, despite the differences in the boson--lepton
coupling between the $\PW$ and the $\PZ$ boson. This is not the case
in the LAB-frame definition.
The LL mode defined in the di-boson CM frame is singled out by the fact that 
both polarization vectors
only depend on the di-boson momentum and on the momentum of a single boson.}
On the contrary, the longitudinal polarization vectors defined in the LAB frame
depend on more physical quantities, resulting in a  {more involved} spin structure.
This qualitative argument indicates that the CM definition of polarization
is more natural than the LAB one for di-boson processes.

A due comment is related to correlations.
The wrong assumption that the spins of the two weak bosons are not correlated would
lead to results which can be far from the correct ones.
When defining polarizations in the CM frame, multiplying the two NLO singly-longitudinal
fractions, one obtains 4.3\%, which is substantially different from the predicted LL
fraction (5.7\%). The difference is larger at LO, where the kinematic is more
constrained due to the absence of additional QCD radiation: the zero-spin-correlation
assumption gives 3.3\% for the LL fraction, whose actual value is 7.9\%.
In the LAB definition, these correlations are smaller: at (N)LO the combination of the
two singly-longitudinal fractions gives 6.4\% (6.5\%), which is not so far from the Monte
Carlo result for the LL fraction, namely 5.7\% (6\%).
The correlations are milder for the mixed states and very small for the TT state,
both in the CM and in the LAB definition.

The rationale is that, in order to correctly model doubly-polarized signals,
all spin-correlations must be properly accounted for. Separating polarizations at the
amplitude level furnishes the safest and most natural strategy in this sense.

\begin{figure*}[h!]
  \centering
  \begin{tabular}{cc}
    \includegraphics[scale=0.42]{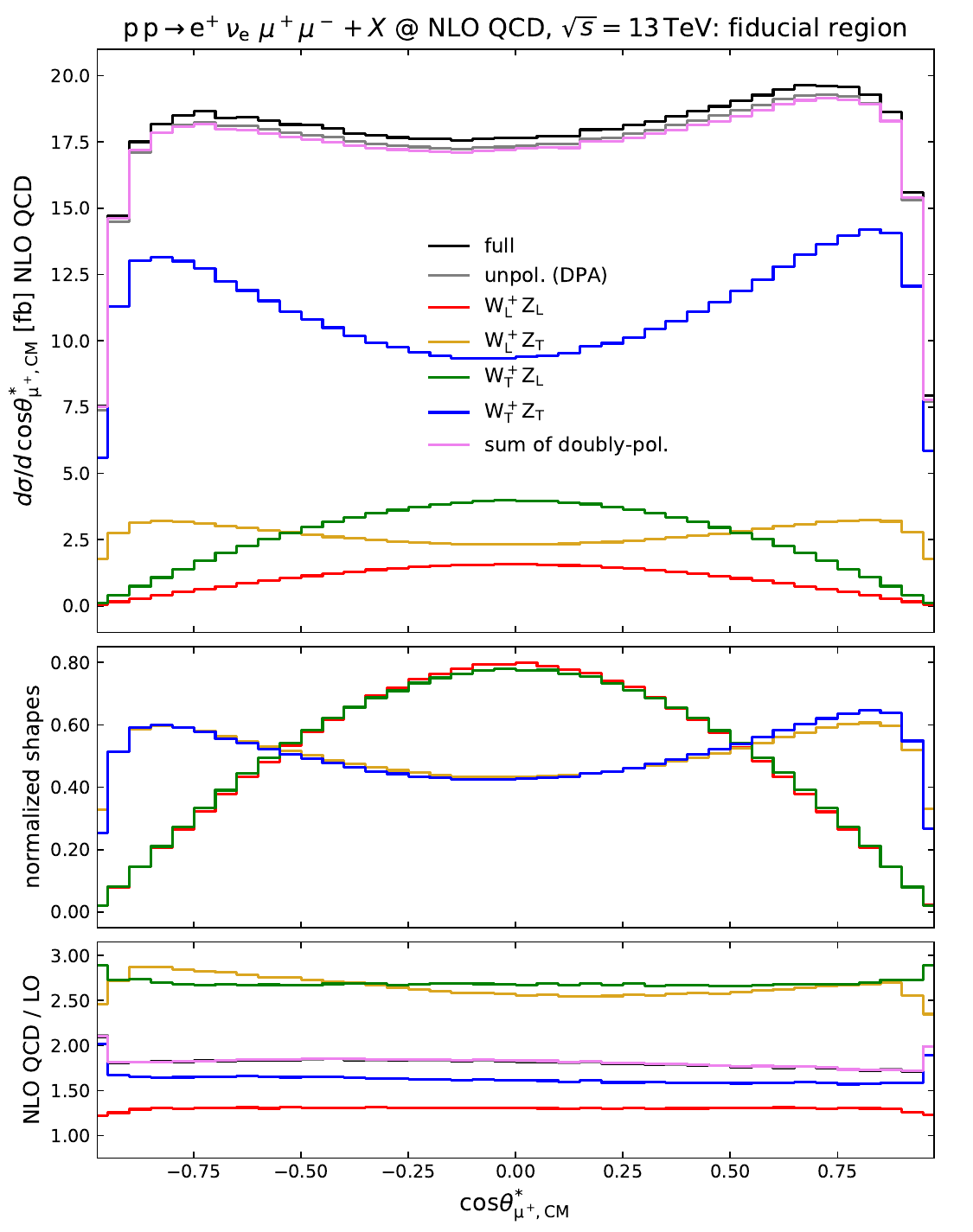} &
    \includegraphics[scale=0.42]{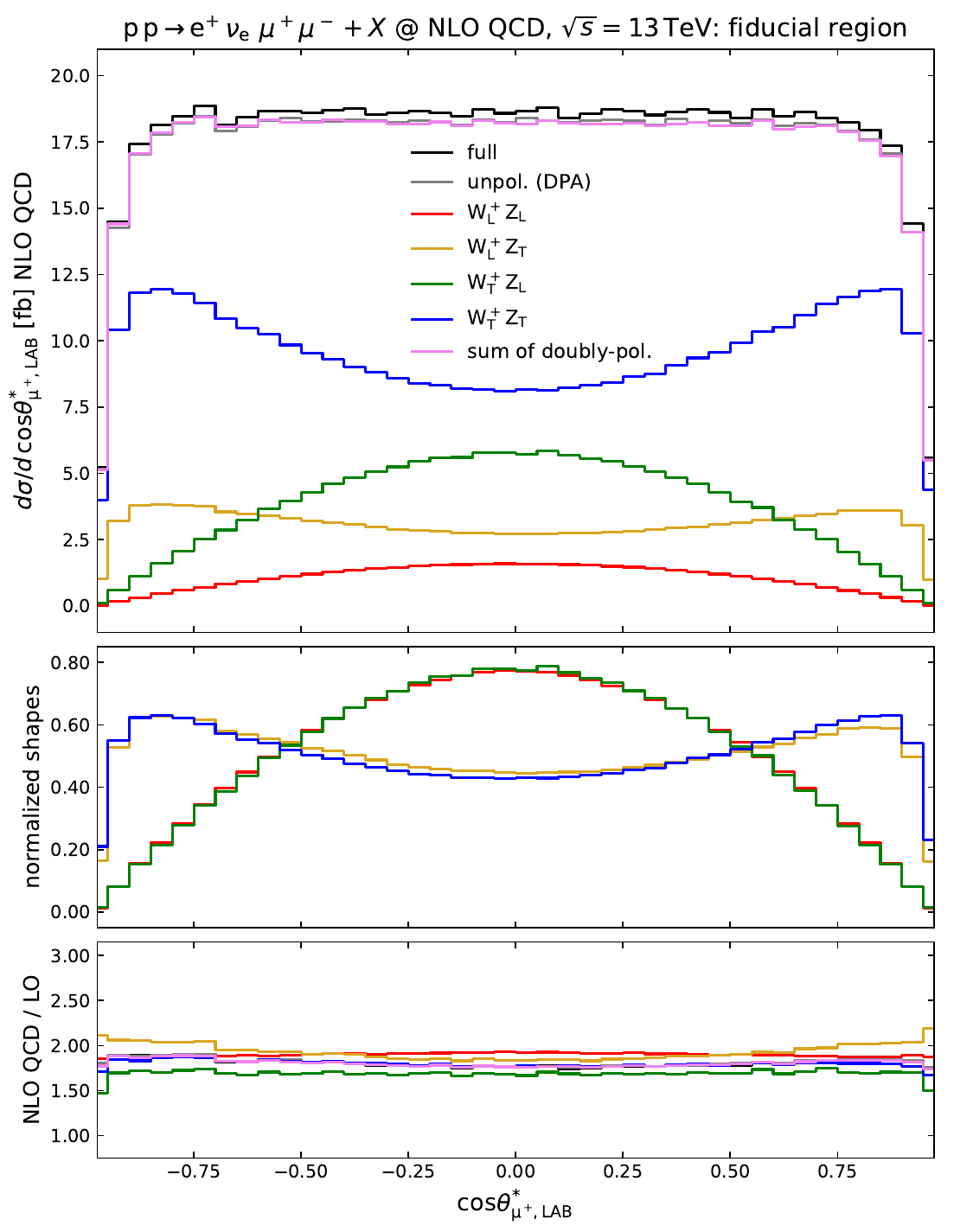} \\
  \end{tabular}
  \caption{Distributions in $\cos\theta^{*}_{\mu^+}$ in the fiducial region for
    polarizations defined in the CM frame (left) and in the LAB frame (right).
   The antimuon angle in the $\PZ$-boson rest frame, $\theta^{*}_{\mu^+}$, is 
    computed with respect to the $\PZ$-boson direction in the CM frame (left)
    and in the LAB frame (right). Doubly-polarized results are shown. From top down:
    NLO QCD differential cross-sections, {NLO QCD} normalized distribution
    shapes (integral equal to 1), $K$-factors (NLO QCD/ LO).}\label{fig:costhetastarmu}
\end{figure*}

\subsection{Distributions}\label{subsec:distrib}
We start this section commenting on the non-resonant background.
At the integrated level, the impact of off-shell effects amounts
to 2\%, and this is also the case in most of the differential
distributions accessible at the LHC. A larger non-resonant
background is found in some suppressed phase-space regions, where
non-resonant diagrams give a sizeable contribution,
and therefore the DPA cannot reproduce accurately the full calculation.
Large effects concern the tails of the distributions for the missing transverse momentum
(10\% at 500\GeV), the invariant mass of the four leptons (20\% at $1.5\TeV$),
and the positron transverse momentum (15\% at $500\GeV$). The largest
non-resonant background is found in the kinematic region where the
transverse mass of the $\Pe^+\nu_{\Pe}$~pair is larger than the $\PW$-boson mass (30\% at 100\GeV, 70\% at 150\GeV).
However, given the limited statistics of Run 2 data, these
regions will be hardly accessible and therefore not really relevant
for the analyses.

Also interferences among polarizations at the differential level reflect
the integrated results in most of the phase-space regions.
Their size never exceeds 2--3\% in all relevant rapidity,
transverse-momentum and angular distributions
of the final-state particles.

\begin{figure*}[h!]
     \centering
     \begin{tabular}{cc}
       \includegraphics[scale=0.42]{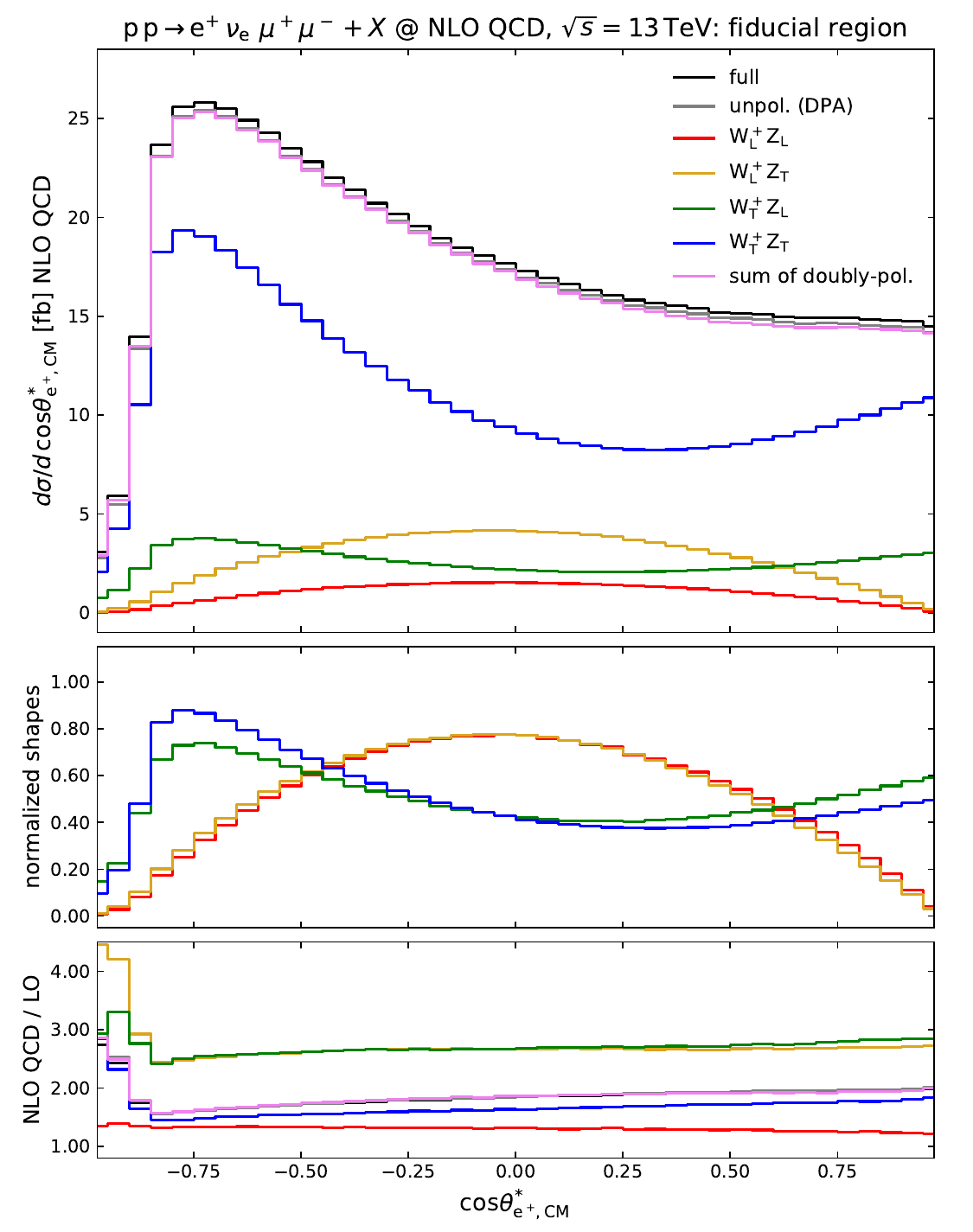} &
       \includegraphics[scale=0.42]{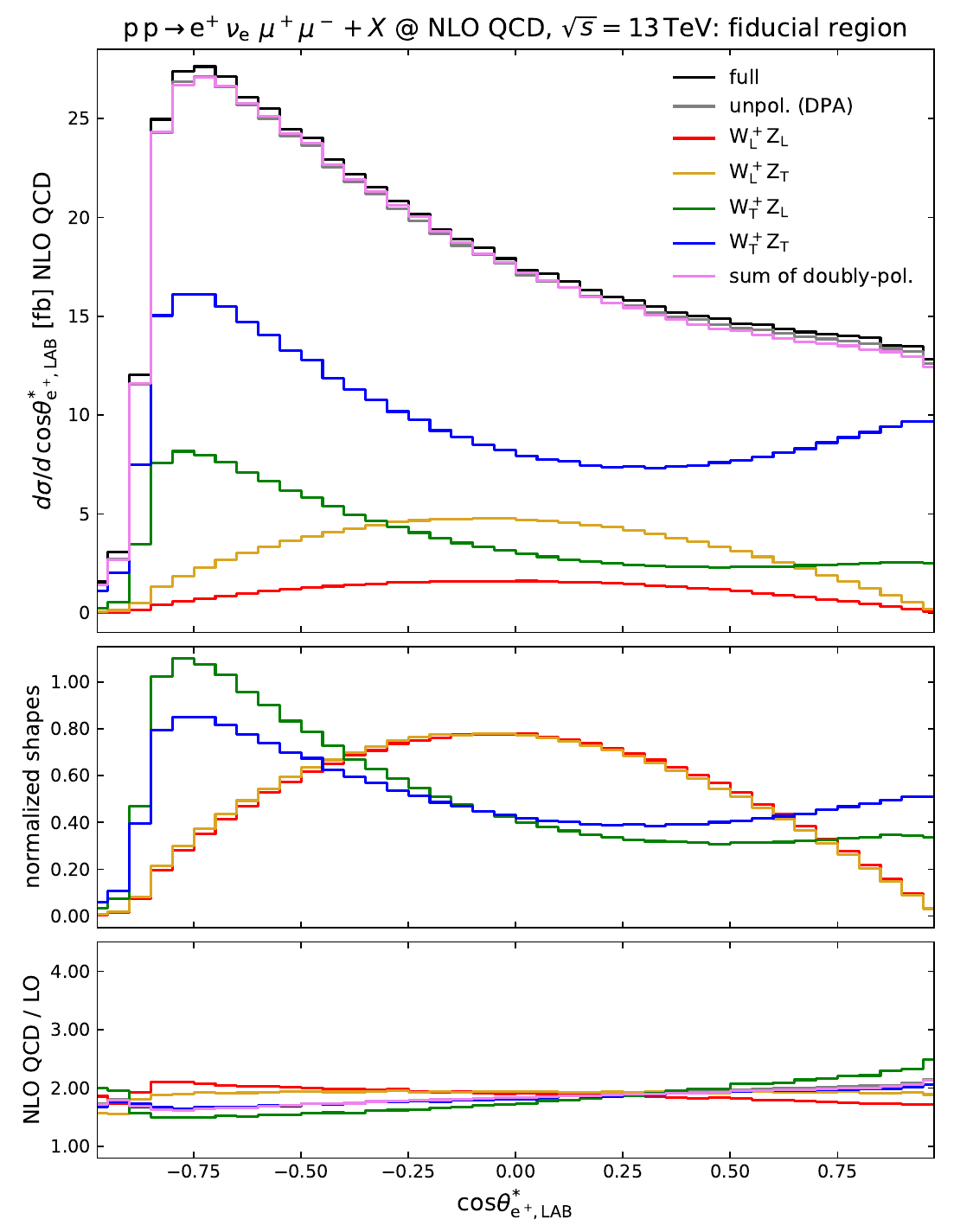} \\
     \end{tabular}
     \caption{Distributions in $\cos\theta^{*}_{\Pe^+}$ in the fiducial region for
       polarizations defined in the CM frame (left) and in the LAB frame (right). 
       The positron angle in the $\PW$-boson rest frame, $\theta^{*}_{\Pe^+}$,
       is computed with respect to the
       $\PW$-boson direction in the CM frame (left) and in the LAB frame (right).
       Same structure as \reffi{fig:costhetastarmu}.}\label{fig:costhetastarpo}
   \end{figure*}

\begin{figure*}[h!]
     \centering
     \begin{tabular}{cc}
       \includegraphics[scale=0.42]{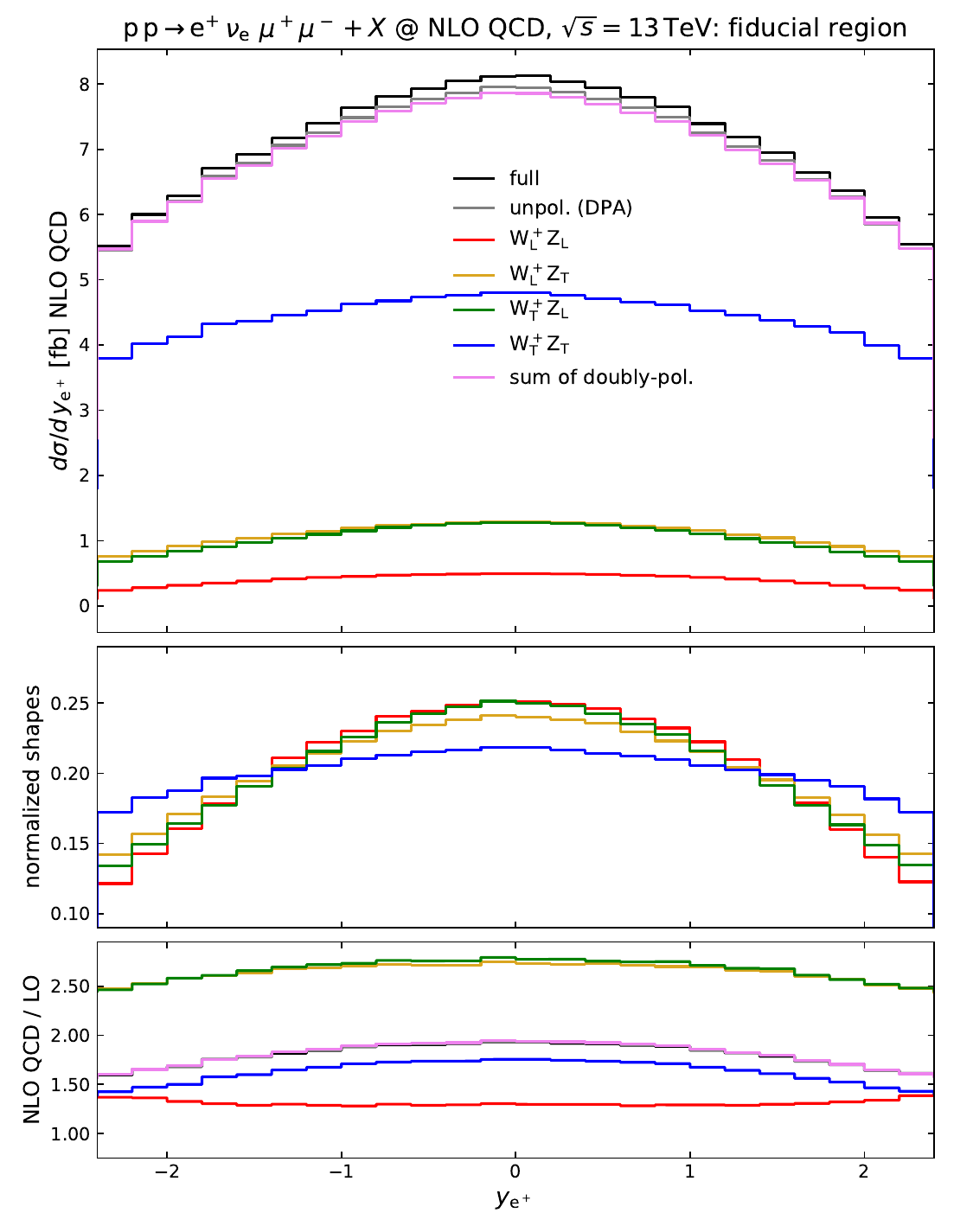} &
       \includegraphics[scale=0.42]{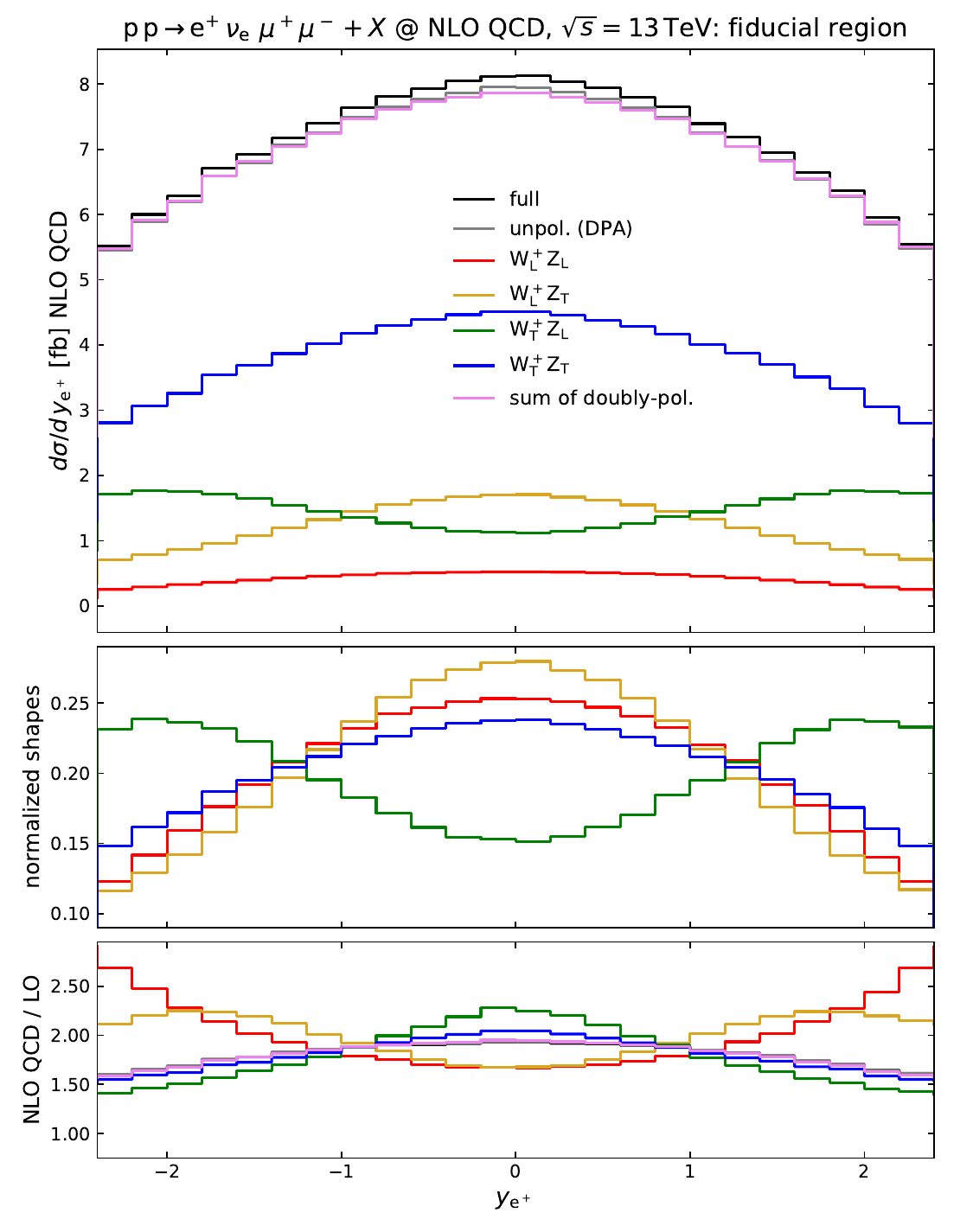}\\
     \end{tabular}
     \caption{Distributions in the positron rapidity in the fiducial region for
       polarizations defined in the CM frame (left) and in the LAB frame (right).
       Same structure as \reffi{fig:costhetastarmu}.
     }\label{fig:rapidity}
   \end{figure*}

\begin{figure*}[h!]
     \centering
     \begin{tabular}{cc}
      \includegraphics[scale=0.42]{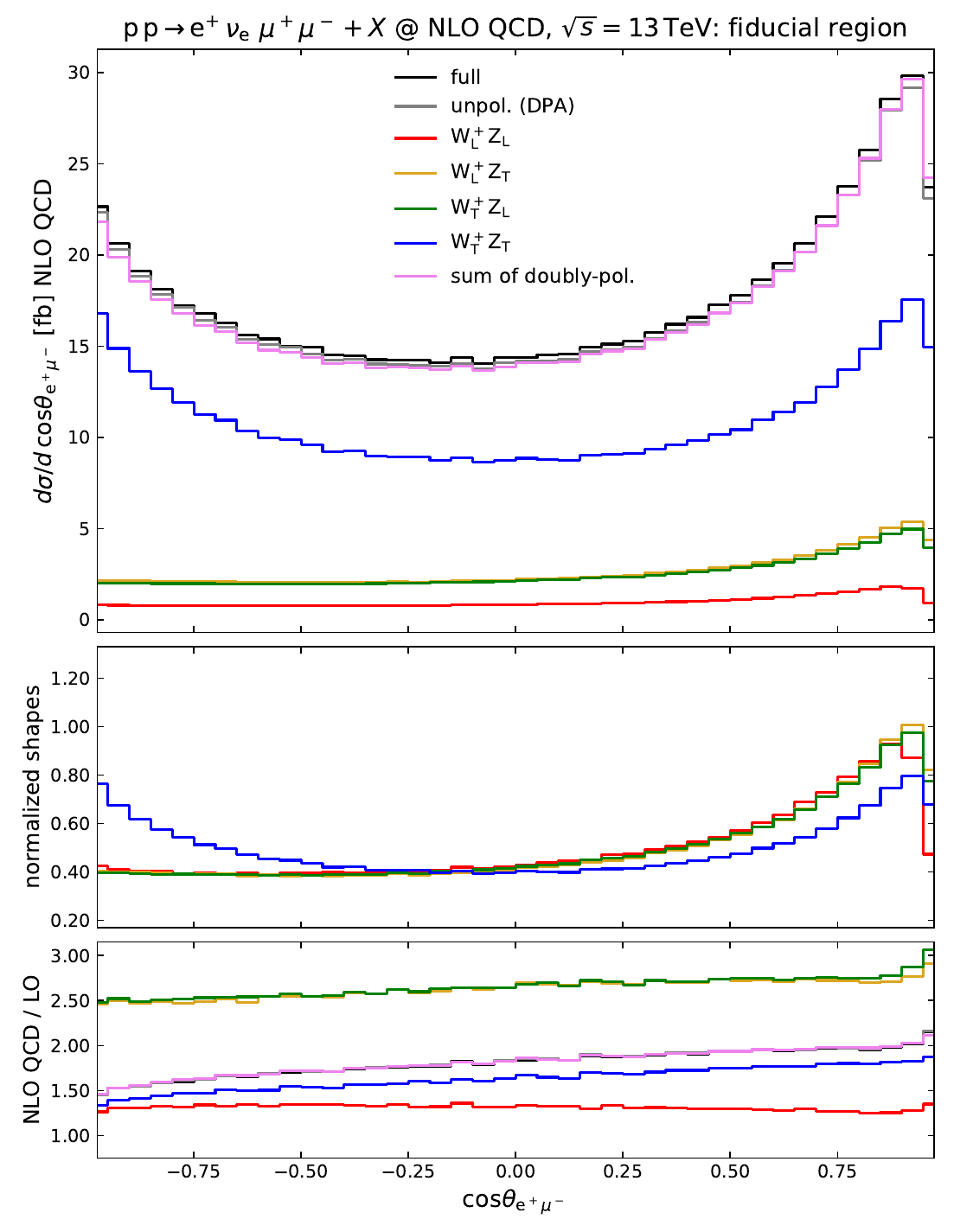} &
      \includegraphics[scale=0.42]{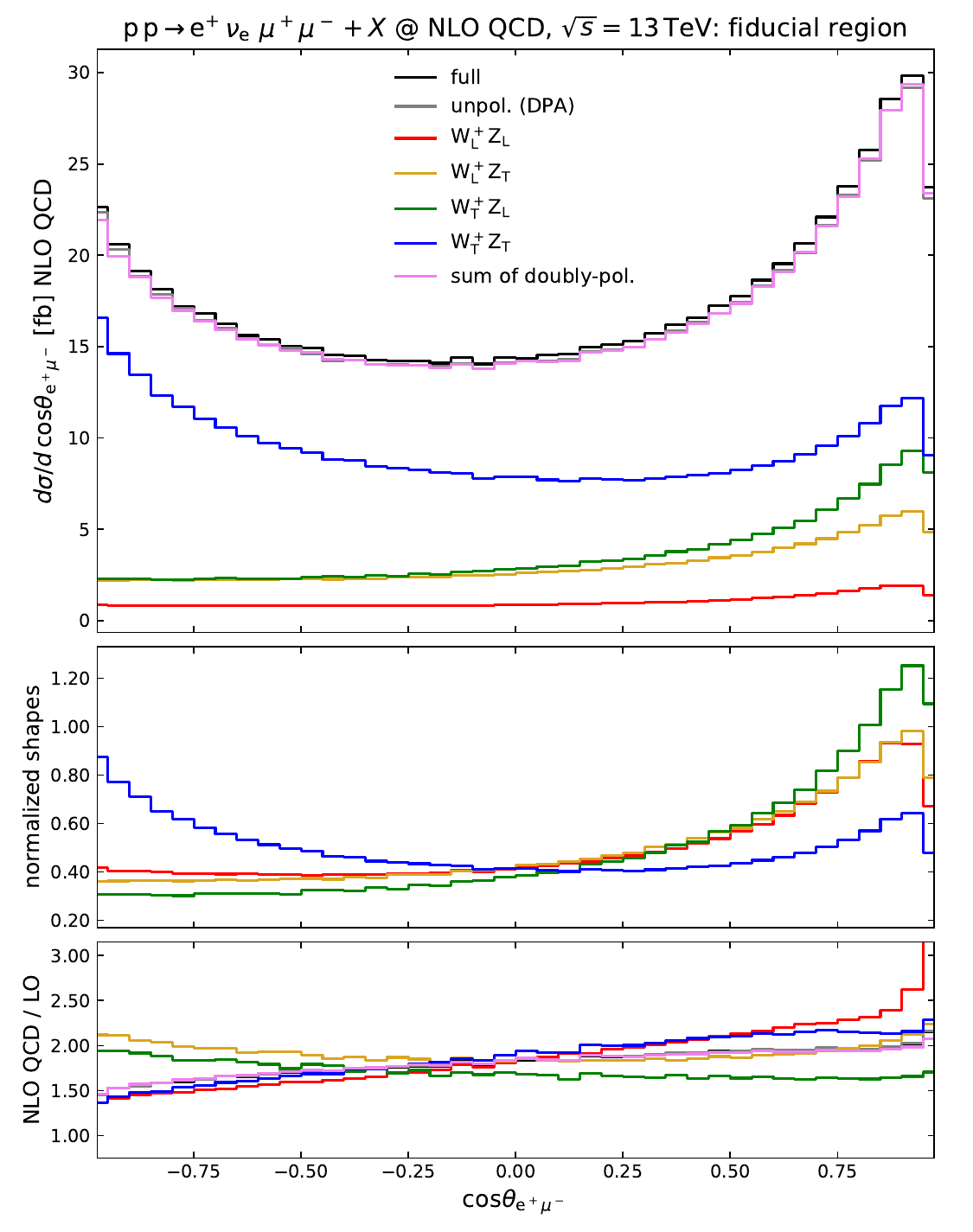}\\
     \end{tabular}
     \caption{Distributions in the cosine of the angle between the positron
       and the muon in the fiducial region for polarizations defined in
       the CM frame (left) and in the LAB frame (right). Same structure as \reffi{fig:costhetastarmu}.
      }\label{fig:coslab}
   \end{figure*}

\begin{figure*}[thb]
       \centering
       \begin{tabular}{cc}
         \includegraphics[scale=0.42]{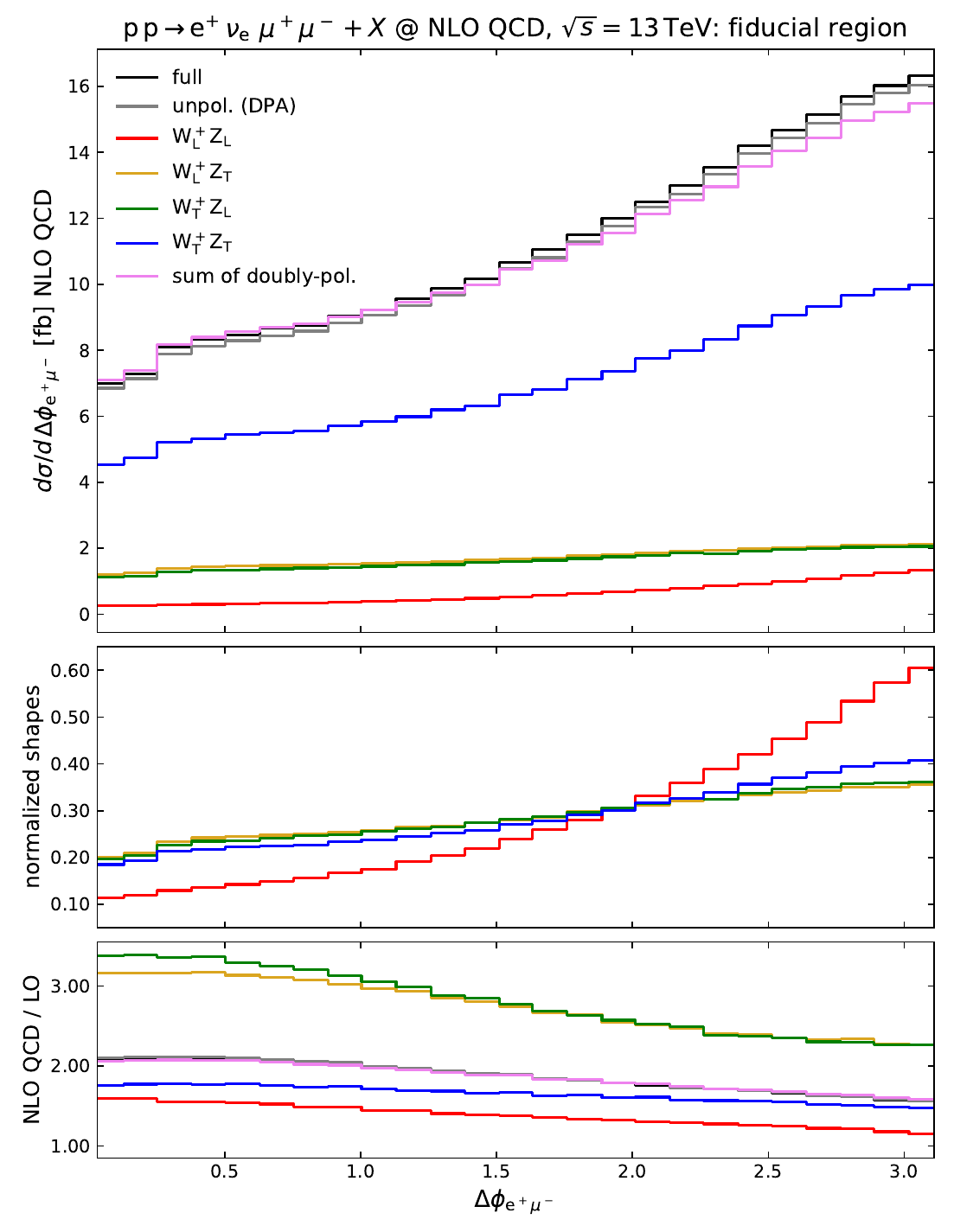} &
         \includegraphics[scale=0.42]{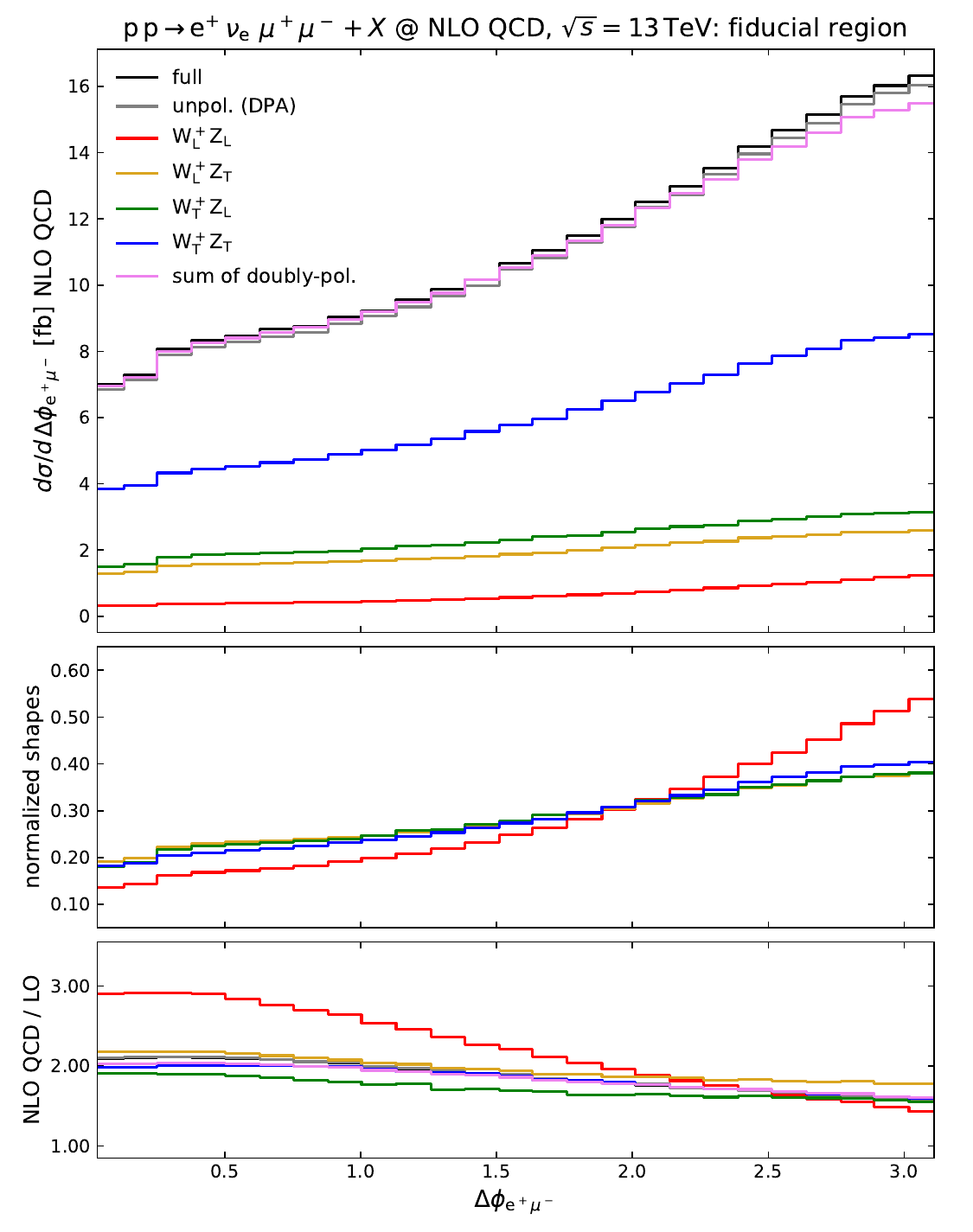}\\
       \end{tabular}
       \caption{Distributions in the azimuthal difference between the
         positron and the muon in the fiducial region for polarizations
         defined in the CM frame (left) and in the LAB frame (right).
         Same structure as \reffi{fig:costhetastarmu}.
       }\label{fig:azimuth}
\end{figure*}

In \reffis{fig:costhetastarmu}--\ref{fig:azimuth} we
show the differential distributions in five significant kinematic variables, for both
the unpolarized and doubly-polarized process.
The non-resonant background is understood as the difference
between the full distribution (black curve) and the DPA unpolarized one (gray curve),
the interference among polarizations as the difference between the DPA
unpolarized one and the sum of doubly-polarized distributions (magenta curve).

The most relevant variables for polarization measurements
are the decay angles of leptons $(\phi_\ell^*,\theta_\ell^*)$ in the corresponding weak-boson rest
frame. At variance with the inclusive setup, in the
presence of lepton cuts a closed analytic formula for $\cos\theta_\ell^*$
distributions is not known, but the distributions for different polarization modes
typically maintain visible shape differences, which make $\cos\theta_\ell^*$
variables the best suited ones for polarization discrimination. Note that if polarization
vectors are defined in the CM (LAB) frame, the angles of the decay lepton $\ell$ of a boson
$V$ should be computed boosting the $\ell$ momentum into the $V$ rest frame from
the CM (LAB) frame. Such variables are not directly accessible at the LHC,
as the $\PW$-boson kinematics (and consequently the one of the di-boson system) is only
known up to neutrino reconstruction, which can be quite inaccurate in describing
some phase-space regions. For the purpose of this work we assume perfect reconstruction
of the neutrino momentum.

In \reffi{fig:costhetastarmu} we consider the distribution in the
antimuon decay angle in the $\PZ$-boson rest frame.  If polarizations
are defined in the LAB frame, this variable is directly observable at
the LHC.  In the CM case, it is subject to the reconstruction of the
di-boson system: we have checked numerically with a neutrino
reconstruction technique used in \citere{Aaboud:2019gxl} that this has
an almost negligible effect on the shape of $\cos\theta^{*}_{\mu^+}$
distributions.  The $K$-factors reflect the integrated results.  Note
that the effect of $p_{\rm T}$ and $\eta$ cuts on $\mu^\pm$ is to cut
down the peaks at $\cos\theta^{*}_{\mu^+} =\pm1$ of the transverse
distributions (TT, LT), while the LL and TL shapes are almost
untouched.  {While the unpolarized distribution shows an
  asymmetry in the {CM} definition, this is washed out when using the
  LAB definition of the decay angle. The asymmetry in the {CM} case
  results from the TT mode, while the distributions are basically
  symmetric for the other polarization modes. For the LAB definition,
  on the other hand, an asymmetry develops for the LT mode which is,
  however, less relevant in view of the much smaller contribution of
  this mode. Up to the differences in the overall cross sections and
  the asymmetries, the description of this variable is very similar in
  the CM and LAB definitions.}
As the normalized shapes in \reffi{fig:costhetastarmu} show,
the $\cos\theta^{*}_{\mu^+}$ variable is only sensitive to the polarization of the $\PZ$~boson.
For a given state of the $\PZ$~boson, the shape is approximately the same for a longitudinal
and a transverse $\PW$~boson. This means that such a variable should be combined with other
discriminating variables to allow for the doubly-polarized extraction via a template-fitting
procedure.

The $\PW^+$-boson polarization state is directly related to the $\cos\theta^{*}_{\Pe^+}$ variable.
It is worth stressing that the neutrino
reconstruction would affect this variable much more than $\cos\theta^{*}_{\mu^+}$, since
it modifies the kinematics of both the di-boson system (needed for the CM definition)
and the $\PW$-boson momentum (needed for both definitions). We have checked that using
the reconstruction of \citere{Aaboud:2019gxl}, the shapes are substantially distorted, leading
to a reduced discrimination power among polarization states.
Assuming perfect neutrino reconstruction, the resulting $\cos\theta^{*}_{\Pe^+}$ distributions
are shown in \reffi{fig:costhetastarpo}.
The transverse and unpolarized distributions are characterized by a depletion at
$\cos\theta^{*}_{\Pe^+}=-1$, but not at $\cos\theta^{*}_{\Pe^+}=1$, since no direct $p_{\rm T}$ or rapidity
constraint is imposed on the neutrino. Analogous situations have been noticed in $\PW^+\PW^-$
production \cite{Denner:2020bcz} and $\PW\PZ$ scattering \cite{Ballestrero:2019qoy}.
In both definitions, the shapes for a longitudinal $\PW$ boson (LL, LT) are made slightly
asymmetric by the lepton cuts, but are almost independent of the $\PZ$-boson polarization.
A different behaviour is found for the two transverse distributions (TL, TT), whose shapes
are mildly sensitive to the $\PZ$-boson polarization. More interestingly, in the CM definition
the TT shape is more peaked at negative values than the TL, while in the LAB one such a difference is
reversed and more sizeable. This means that, depending on the $\PZ$ polarization, there is
a different left--right polarization balance of the \PW~boson.
The $K$-factors are roughly equal to those of the integrated
cross-sections, up to some deviation in the
anti-collinear regime for the CM-frame transverse modes. In both definitions, the $K$-factors
are less flat than those observed in $\cos\theta^*_{\mu^+}$ distributions.
The different behaviour of $\cos\theta^*_{\Pe^+}$ distributions with respect to the
$\cos\theta^*_{\mu^+}$ ones can be traced back to the different nature of the cuts
imposed on the $\PW$-decay leptons (asymmetric, stronger $p_{\rm T}$ cut, transverse-mass
constraint) and of those imposed on the $\PZ$-boson decay products (symmetric,
weaker $p_{\rm T}$ cut, invariant-mass constraint).

A satisfactory fit of LHC data with SM templates can only 
be achieved with the combination of several kinematic observables which are best
suited to discriminate among doubly-polarized signals.
We have generated NLO QCD accurate SM templates for many observables. In general,
the transverse-momentum distributions (for $\Pe^+, \mu^\pm, \PZ$ and missing energy)
do not show sizeable differences among polarized signals. A mild discrimination
power, in particular for the LL mode, is found in the rapidity
separation between the positron and the muon, 
both in the CM and in the LAB definition.

Noticeable differences between the LAB and CM definitions can be seen in the
positron-rapidity distributions, which are shown in \reffi{fig:rapidity}.
This variable shows almost negligible interferences and
a non-resonant background in line with the integrated result.
The relative NLO QCD corrections to the LL signal become maximal near
the cut
$y_{\Pe^+} = \pm 2.5$ in both definitions, but the size is much larger
in the LAB-frame definition than in the CM one.
All other polarized signals based on the CM definition have roughly the same
differential $K$-factors, with QCD corrections that are maximal
at zero rapidity.
Differently from TT and TL, the LT signal in the LAB definition is
mostly enhanced by QCD corrections near $y_{\Pe^+}=\pm 2$.
In the CM definition, all polarized signals have a maximum at
zero rapidity, though with somewhat different shapes. In
particular, the TT signal is flatter than all the other ones.
Much more sizeable shape differences characterize the signals
in the LAB definition. The TL signal has two maxima around
$y_{\Pe^+}=\pm 2$, and a minumum at zero rapidity, where all
other doubly-polarized modes have a maximum. The LL, LT, and
TT modes have similar normalized shapes, but slightly
different peak widths.

{
  The marked difference between the TL and LT distributions
  in the LAB-frame definition, which is even more evident at
  LO, can be traced back to the boost that relates the LAB
  and the CM reference frames.
  If the polarizations are defined in the CM frame, all polarized
  rapidity distributions follow similar patterns, as might be expected
  from the more universal character of this definition.
  However, due to different PDF densities (the u-quark PDF peaks at
  larger energies than the $\bar{\rm d}$-quark one) connecting the CM
  frame to the LAB frame involves a boost preferentially in the
  direction of the u quark. Moreover, in the CM frame the $\PW^+$
  boson is preferably produced in the hemisphere determined by the
  u-quark direction. As a consequence, the boost to the LAB frame
  typically accelerates the $\PW^+$ boson, while it decelerates the
  \PZ~boson.  Upon inspecting the explicit form of the polarization
  vectors, it can be seen that a boost that increases the energy
  changes the polarization vectors only mildly, while a boost that
  reduces the energy yields more sizeable changes in polarization.
  In extreme cases the boost results in a very small momentum
  along the $z$ axis or even reverses this component. Thus, in events
  with a large $y_{\PW^+}$ the polarization of the $\PW^+$ boson is hardly
  changed by the boost to the LAB frame, while the one for the $\PZ$
  changes relatively often.  On the other hand, in events with a small
  $y_{\PW^+}$ the polarization of the $\PW^+$ is likely to change, while
  the one of the $\PZ$ is pretty stable. Since the $\PW$ rapidity is
  correlated to the positron rapidity, this explains the enhancement
  of the $\PW_\rT^+\PZ_\rL$ mode for large $y_{\Pe^+}$ and the
  enhancement of the $\PW_\rL^+\PZ_\rT$ mode for small $y_{\Pe^+}$ via
  migration from the dominant $\PW_\rT^+\PZ_\rT$ mode.

  %
}

Even though the results for the positron rapidity suggest that investigating
polarizations defined in the LAB frame provides more discrimination power
than in the CM frame, we stress that the measured quantities
are different in the two cases and cannot be related to each other.
Thus, two separate analyses for polarizations defined in the CM frame and in
the LAB frame are useful.  

We have investigated more rapidity observables (for $\PZ$, $\mu^\pm$)
which feature similar or smaller discrimination power than $y_{\Pe^+}$.

Another interesting variable accessible at the LHC
is the cosine of the angle $\theta_{\Pe^+\mu^-}$ between the positron
and the muon, which is presented in \reffi{fig:coslab}.
In the fiducial region, the two leptons tend to be mostly collinear
in the unpolarized case, but different behaviours are found for the
polarized modes. The relative QCD corrections are rather
flat for the LL signal defined in the CM frame, but increasing
with $\cos\theta_{\Pe^+\mu^-}$ for the corresponding signal defined in the LAB frame.
The $K$-factor for the TT mode is monotonically increasing and similar
in the two definitions. A similar pattern is found for TL and LT defined in
the CM frame, up to the overall shift already commented at the integrated
level. The mixed contributions show decreasing $K$-factors in the
LAB definition.
In the CM definition, the normalized TT distribution is almost symmetric, while
modes including a longitudinal boson are rather flat for negative
values and have a maximum in the collinear regime. Very small shape
differences characterize the LL, LT and TL distributions. Only the TT
signal can be clearly disentangled.
For polarizations defined in the LAB frame, the TT distribution favours
anticollinear configurations, while the LL and LT modes show similar
behaviours as for the CM definitions. The TL distribution has a peak in the collinear
region that is more marked than the one of LL and LT modes.
{The more sizeable differences among different polarization
  states in the LAB frame for large $\cos\theta_{\Pe^+\mu^-}$ can
  again be explained by effects of the boost from the CM frame to the
  LAB one and its effect on the polarization of the weak bosons.  A
  small angle between positron and muon is correlated to a small angle
  between the $\PW^+$ and the $\PZ$ boson. This requires the reversal
  of the $z$ component of one of the vector bosons, mostly of the
  $\PZ$ boson, by the boost and thus explains the observed sizeable reshuffling
  between polarized cross-sections for large  $\cos\theta_{\Pe^+\mu^-}$.
}

In \reffi{fig:azimuth} we show the distribution in the azimuthal difference
between the positron and the muon.
This observable is characterized by
negative interferences for $\Delta\phi_{\Pe^+\mu^-}\lesssim \pi/2$ and
positive ones for
$\Delta\phi_{\Pe^+\mu^-}\gtrsim \pi/2$ in both the CM- and the LAB-frame polarization
definitions. Although the interferences are similar in shape to those found
in $\PW^+\PW^-$ production \cite{Denner:2020bcz}, their size ($\pm 3.5\%$ for
$\Delta\phi_{\Pe^+\mu^-}\!=0,\pi$) is much smaller than the $\pm 50\%$ effect
found there.
The polarized distributions become maximal at $\Delta\phi_{\Pe^+\mu^-}=\pi$
in both polarization definitions, and all $K$-factors are decreasing
with growing $\Delta\phi_{\Pe^+\mu^-}$. The
$K$-factor of the LL mode is flatter in the CM definition, while the one of the
mixed contributions is flatter in the LAB one. All polarization states with at
least one transverse boson exhibit roughly the same normalized distribution
shape, which is also very similar in the two definitions of polarization
vectors. A significantly different shape is found in the LL
polarization state both in the LAB and even stronger in the CM definition. This
makes this azimuthal separation a good candidate variable to improve the
sensitivity to the purely longitudinal cross-section. 

As a last comment of this section, we stress that the differential
distributions analyzed in this paper do not provide an exhaustive
list of physical quantities relevant for the extraction of doubly-polarized
signals out of $\PW\PZ$ data, but are representative for observables
that are best suited for polarization discrimination (in the CM and/or
in the LAB definition), and that highlight relevant differences
between the two polarization definitions.

\section{Conclusions}\label{concl}
In this paper we have presented integrated and differential
cross-sections for the $\PW\PZ$ production at the LHC with both bosons
polarized.  The results feature NLO QCD accuracy and have been
obtained for realistic selection cuts.

The non-resonant background, estimated as the effects missing
in the double-pole-approximated calculation, are at the 2\% level for the integrated
cross-sections. Slightly larger effects are found in the tails of some distributions,
which are, however, out of reach with the statistics of the Run 2 data.

We have investigated polarized signals defining polarization vectors
both in the di-boson centre-of-mass (CM) and in the laboratory (LAB) system.
In both cases the interferences among different polarization modes are very small
(of order 1\%).
The integrated and differential cross-sections for a given polarization state
obtained with the two definitions are rather different, in particular, for the doubly-polarized signals.
We have analyzed the impact of NLO QCD corrections, the relative fractions for the various
polarization states, and the normalized shapes for several kinematic distributions.

A small number of observables are characterized by a noticeable discrimination
power among polarizations. Some of them are more sensitive to polarizations
defined in the CM frame, some to those defined in the LAB frame. This suggests that
both definitions should be taken into account in the upcoming experimental analyses.

\section*{Acknowledgements}
We thank Lucia Di Ciaccio, Corinne Goy, Francesco Costanza and Emmanuel Sauvan
for useful discussions.
The authors are supported by the German Federal Ministry for Education and Research (BMBF) under contract 
no.~05H18WWCA1.

 \bibliographystyle{elsarticle-num} 
 \biboptions{numbers,sort&compress}
 \bibliography{polwz}

\end{document}

%% file: macros.tex
\def\refeq#1{\mbox{(\ref{#1})}}
\def\reffi#1{\mbox{Figure~\ref{#1}}}
\def\reffis#1{\mbox{Figures~\ref{#1}}}

\def\citere#1{\mbox{Ref.~\cite{#1}}}
\def\citeres#1{\mbox{Refs.~\cite{#1}}}

\newcommand{\newc}{\newcommand}
\newc{\beq}{\begin{equation}}
\newc{\eeq}{\end{equation}}
\newc{\bit}{\begin{itemize}}
\newc{\eit}{\end{itemize}}
\newc{\ben}{\begin{enumerate}}
\newc{\een}{\end{enumerate}}
\newc{\bce}{\begin{center}}
\newc{\ece}{\end{center}}
\newc{\bfi}{\begin{figure}}
\newc{\efi}{\end{figure}}

\newcommand{\rT}{{\mathrm{T}}}
\newcommand{\rR}{{\mathrm{R}}}
\newcommand{\rL}{{\mathrm{L}}}
\newcommand{\rF}{{\mathrm{F}}}
\newcommand{\rU}{{\mathrm{U}}}

\newcommand{\GeV}{\ensuremath{\,\text{GeV}}\xspace}
\newcommand{\TeV}{\ensuremath{\,\text{TeV}}\xspace}

\newcommand{\Pp}{\ensuremath{\text{p}}}
\newcommand{\Pe}{\ensuremath{\text{e}}\xspace}

\newcommand{\Pg}{\ensuremath{\text{g}}}

\newcommand{\PW}{\ensuremath{\text{W}}\xspace}
\newcommand{\PZ}{\ensuremath{\text{Z}}\xspace}

\newcommand{\pt}[1]{p_{\rT,{#1}}}
\newcommand{\mt}[1]{M_{\rT,{#1}}}

\newcommand{\MW}{\ensuremath{M_\PW}\xspace}

\newcommand{\MZ}{\ensuremath{M_\PZ}\xspace}

\newcommand{\GZ}{\ensuremath{\Gamma_\PZ}\xspace}

\newcommand{\GW}{\ensuremath{\Gamma_\PW}\xspace}

\newcommand{\GF}{\ensuremath{G_\mu}}
\newcommand{\alphas}{\ensuremath{\alpha_\text{s}}\xspace}

\newcommand{\recola}{{\sc Recola}\xspace}

\newcommand{\mocanlo}{{\sc MoCaNLO}\xspace}
\newcommand{\collier}{{\sc Collier}\xspace}

\newcommand{\madgraph}{{\sc MadGraph}\xspace}

\newcolumntype{.}{D{.}{.}{-1}}
\newcolumntype{d}[1]{D{.}{.}{#1}}
\colorlet{tableoverheadcolor}{gray!37.5}
\colorlet{tableheadcolor}{gray!25}
\colorlet{tablerowcolor}{gray!12.5}

\def\draftdate{\relax}
\def\mda{\relax}
\def\mua{\relax}
\def\mla{\relax}
\def\draft{
\def\thtystars{******************************}
\def\sixtystars{\thtystars\thtystars}
\typeout{}
\typeout{\sixtystars**}
\typeout{* Draft mode!
         For final version remove \protect\draft\space in source file *}
\typeout{\sixtystars**}
\typeout{}
\def\draftdate{\today}
\def\mua{\marginpar[\boldmath\hfil$\uparrow$]%
                   {\boldmath$\uparrow$\hfil}\color{black}%
                    \typeout{marginpar: $\uparrow$}\ignorespaces}
\def\mda{\color{red}\marginpar[\boldmath\hfil$\downarrow$]%
                   {\boldmath$\downarrow$\hfil}%
                    \typeout{marginpar: $\downarrow$}\ignorespaces}
\def\mla{\marginpar[\boldmath\hfil$\rightarrow$]%
                   {\boldmath$\leftarrow $\hfil}%
                    \typeout{marginpar: $\leftrightarrow$}\ignorespaces}
\def\Mua{\marginpar[\boldmath\hfil$\Uparrow$]%
                   {\boldmath$\Uparrow$\hfil}\color{black}%
                    \typeout{marginpar: $\uparrow$}\ignorespaces}
\def\Mda{\color{red}\marginpar[\boldmath\hfil$\Downarrow$]%
                   {\boldmath$\Downarrow$\hfil}%
                    \typeout{marginpar: $\downarrow$}\ignorespaces}
\def\Mla{\marginpar[\boldmath\hfil\textcolor{red}{$\Rightarrow$}]%
                   {\boldmath\textcolor{red}{$\Leftarrow $}\hfil}%
                    \typeout{marginpar: $\leftrightarrow$}\ignorespaces}
\overfullrule 5pt
\oddsidemargin 15mm
\marginparwidth 29mm
}
